\documentclass[conference,compsoc]{IEEEtran}
\usepackage[usenames,dvipsnames,svgnames,x11names]{xcolor}

\newcommand{\ysnoted}[1]{} 

\newcommand{\defn}[2]{\item \textbf{#1}:~#2}
\usepackage[normalem]{ulem} 

\usepackage[nocompress]{cite}
\usepackage{lipsum}
\let\OLDthebibliography\thebibliography
\renewcommand\thebibliography[1]{
  \OLDthebibliography{#1}
  \setlength{\parskip}{0pt}
  \setlength{\itemsep}{0pt plus 0.3ex}
}

\usepackage{listings}

\lstdefinelanguage{XML}
{
basicstyle=\ttfamily\footnotesize,
  morestring=[b]",
  moredelim=[s][\bfseries\color{Maroon}]{<}{\ },
  moredelim=[s][\bfseries\color{Maroon}]{</}{>},
  moredelim=[l][\bfseries\color{Maroon}]{/>},
  moredelim=[l][\bfseries\color{Maroon}]{>},
  morecomment=[s]{<?}{?>},
  morecomment=[s]{<!--}{-->},
  commentstyle=\color{gray},
  stringstyle=\color{blue},
  identifierstyle=\color{red}
}
\usepackage{moreverb}

\usepackage[nounderscore]{syntax}

\usepackage[pdftex]{graphicx}
\graphicspath{{./figures/}}
\DeclareGraphicsExtensions{.pdf}
\usepackage{subfig} 
\usepackage[cmex10]{amsmath}
\usepackage{amssymb}
\usepackage{mathtools}
\usepackage{amsthm}
\usepackage{amsfonts}

\usepackage{algorithmicx}
\usepackage{algpseudocode}
\usepackage[ruled]{algorithm}
\definecolor{light-gray}{gray}{0.75}
\algrenewcommand{\algorithmiccomment}[1]{\hskip3em{{\footnotesize \textcolor{light-gray}{$\blacktriangleright$}}} #1}
\usepackage{multirow} 
\usepackage{rotating} 
\usepackage{booktabs} 
\usepackage{colortbl} 
\usepackage{tablefootnote} 
\usepackage{array}
\newcolumntype{R}[1]{>{\raggedleft\let\newline\\\arraybackslash\hspace{0pt}}m{#1}}

\usepackage[pdftex,colorlinks=true,urlcolor=blue,citecolor=blue]{hyperref}

\usepackage{xspace}

\usepackage{enumitem}

\usepackage[english]{babel}
\usepackage{blindtext}

\begin{document}
%
\title{Toward Reliable and Rapid Elasticity for Streaming Dataflows on Clouds}

\author{\IEEEauthorblockN{Anshu Shukla and Yogesh Simmhan}
\IEEEauthorblockA{Department of Computational and Data Sciences\\
Indian Institute of Science, Bangalore 560012, India\\
Email: shukla@grads.cds.iisc.ac.in, simmhan@iisc.ac.in}
}


%


\maketitle

\begin{abstract}
The pervasive availability of streaming data is driving interest in distributed Fast Data platforms for streaming applications. Such latency-sensitive applications need to respond to dynamism in the input rates and task behavior using scale-in and -out on elastic Cloud resources. Platforms like \emph{Apache Storm} do not provide robust capabilities for responding to such dynamism and for rapid task migration across VMs. We propose several dataflow checkpoint and migration approaches that allow a running streaming dataflow to migrate, without any loss of in-flight messages or their internal tasks states, while reducing the time to recover and stabilize. We implement and evaluate these migration strategies on Apache Storm using micro and application dataflows for scaling in and out on up to $2-21$ Azure VMs. Our results show that we can migrate dataflows of large sizes within $50~sec$, in comparison to Storm's default approach that takes over $100~sec$. We also find that our approaches stabilize the application much earlier and there is no failure and re-processing of messages.

\end{abstract}


%
\IEEEpeerreviewmaketitle

\section{Introduction}
\label{sec:intro}

The rapid growth of observational and streaming data on physical systems and online services is driving the need for applications that operate on these streams in near real-time. Traditional stream sources, like micro-blogs and social networks, financial transactions and web logs, are being complemented by sensor observations from \emph{Internet of Things (IoT)} domains, such as smart power grids, personal fitness devices, and autonomous vehicles. Such streams are analyzed for live visualization in dashboards or to perform online decision making, such as to trigger load curtailment in power grids or to alert users to suspicious transactions~\cite{simmhan:cise:2013}.

Stream processing applications need to operate with low latency to respond rapidly to evolving situations, and need to scale with the number and the rate of the streams. Fast data platforms, also called \emph{Distributed Stream Processing Systems (DSPS)}, offer composition environment to design such applications as a dataflow graph of user-defined tasks, and execute them on distributed resources such as commodity clusters and Clouds. 
Frameworks such as \emph{Apache Storm, Flink} and \emph{Spark Streaming} are popular to support this velocity dimension of Big Data~\cite{laney:gartner:2001}.

Streaming applications are sensitive to \emph{dynamism} -- be it changes in the input stream due to sampling, in the tasks' behavior and resource requirements, or the Virtual Machine's (VM) performance due to multi-tenancy. Such variations can cause the dataflow's performance (e.g., latency, supported throughput) to be affected, and violate the application's Quality of Service (QoS) requirement. While fast data platforms are designed to scale, 
they are less responsive to such dynamism and 
have limited ability to change the dataflow's schedule at runtime. But this feature is essential to leverage the elasticity offered by Cloud VMs to respond to changing situations, and to efficiently utilize pay-as-you-go Cloud resources, say, by consolidating tasks from many under-utilized VMs to fewer well-utilized VMs.

E.g., Apache Storm, a popular open-source stream processing platform from Twitter, uses the R-Storm scheduler for resource-aware scheduling when the dataflow is submitted~\cite{peng:middleware:2015}. However, any changes in the stream, dataflow or VMs' performance needs the user's intervention to ``rebalance'' the placement of tasks onto the same or a different set of VMs. 
A key challenge is to enact this rebalance such that: (1) there is no loss of messages or task states, and (2) it is done rapidly with minimal turn around time~\footnote{A related but separate problem is to determine the new resource allocation for the dataflow (number and sizes of VMs) and the new mapping of its tasks onto the VMs. This is outside the scope of this paper, but has been examined elsewhere~\cite{shukla:scheduler:2017}. Having a new schedule is a precursor to the dynamic enactment of the schedule, which we target in the current paper.}. The former ensures consistency and reliability, while the latter is important for mission-critical applications that cannot suffer prolonged down-time during this rebalance. 

Both of these are lacking in contemporary DSPS. They require the dataflow to be halted before rebalancing it, causing message and task state loss in the process. Platforms like Storm and Flink have robust mechanisms for replaying lost messages, and for regularly checkpointing the state of tasks in the dataflow. These can be leveraged to ensure reliability after rebalance. However, these fault-tolerance methods tend to be disruptive -- message or machine loss are less frequent, while planned rebalance can happen more frequently. Prior research like ElasticStream~\cite{ishii:cloud:2011} plan to minimize Cloud costs by dynamically adjusting the resources required with input rates, while Stela~\cite{xu:ic2e:2016} does on-demand scaling for Storm while optimizing the  throughput and limiting application interruption. Others~\cite{gedik:tpds:2014} perform incremental migration to maintain states while scaling, minimizing the amount of state transfer between hosts. But none of these address message reliabilty and state handling during the migration.

In this paper, we propose mechanisms to dynamically enact the rescheduling and migration of tasks in a streaming dataflow from one set of VMs to another, reliably and rapidly. Specifically, we make the following contributions:
\begin{enumerate}
\item We discuss current rebalance capabilities of Storm, as an exemplar DSPS, and motivate the need for better approaches to migration of streaming dataflows (\S~\ref{sec:background}).
\item We propose two novel migration strategies, \emph{Drain-Checkpoint-Restore (DCR)} and \emph{Capture-Checkpoint-Resume (CCR)} (\S~\ref{sec:approach}), besides a baseline approach, that are implemented on Storm.
\item We introduce \emph{metrics} to evaluate these strategies (\S~\ref{sec:metrics}), and \emph{evaluate} the performance of the approaches for realistic dataflows on Storm within Azure Cloud (\S~\ref{sec:results}).
\end{enumerate}

Finally, related work is reviewed in~\S~\ref{sec:related} and our conclusions and future work presented in~\S~\ref{sec:conclusions}.

\section{Background and Motivation}
\label{sec:background}
\begin{figure}[t]
	\centering
        \includegraphics[width=0.95\columnwidth]{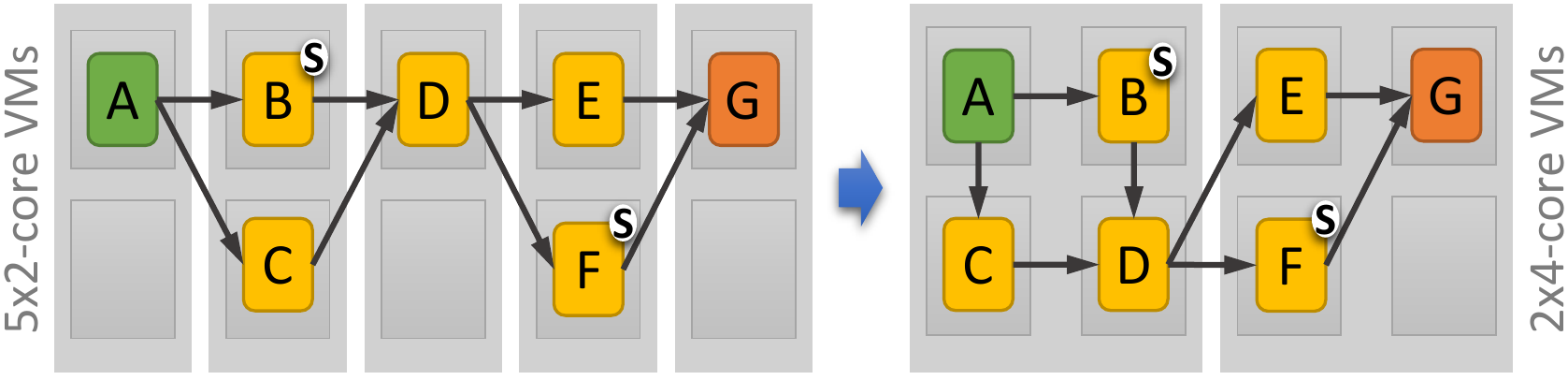}
	\caption{Example of migration of \emph{Star} dataflow from $5\times 2$-core VMs to $2\times 4$-core VMs. `\textbf{s}' indicates a stateful task.}
	\label{fig:migrate}
\vspace{-0.15in}
\end{figure}

In this section, we motivate the need for dynamic migration of streaming applications across elastic Cloud resources. We also provide background on Apache Storm's reliability and rebalancing capabilities, as a representative DSPS, highlighting the gaps in the existing capabilities.

\textbf{Streaming dataflow applications} are composed as a directed graph of tasks, with event stream(s) initiated at one or more \emph{source} tasks from external sources, processed and streamed through additional tasks, and finally terminating in one or more \emph{sink} tasks that may persist or publish the output stream(s). Fig.~\ref{fig:migrate} shows a \emph{Star} dataflow with $7$ tasks, $A-G$, with one source (green, $A$) and one sink task (red, $G$)~\cite{peng:middleware:2015}. These tasks are active all times, and the user-logic in the task is executed for each event as it arrives on the input stream. 
Tasks can be \emph{stateful}, where its execution depends on and can update a local in-memory state. E.g., $B$ and $F$ are stateful tasks, and may maintain, say, a count of events seen or a windows of events for aggregation.

DSPS like Storm coordinate the \textbf{resource allocation, placement and execution} of the dataflow on distributed resources. Typically, VMs available within the shared DSPS cluster are divided into logical \emph{resource slots}, and tasks of the dataflow are placed in these slots at deployment time. Multiple instances of a tasks may also be present, based on the degree of data parallelism required, and can share the same slot. Fig.~\ref{fig:migrate}(left) shows the $7$ tasks of the Star dataflow with one instance each being placed on $7$ slots spread across $5$ VMs, each having $2$ slots of 1-core each.

\textbf{Dynamism} in the input event rates, resources consumed by the tasks, QoS requirements of the application, or the performance of the VMs can cause the initial resource allocation (number and size of VMs) and placement (mapping of tasks to slots) to become sub-optimal. Then, some or all tasks in the dataflow will need to be migrated to other slots in an independent or an intersecting set of VMs. This includes \emph{consolidation} to fewer VMs to reduce costs and improve locality/latency, \emph{scale-out} to more VMs to respond to increased resource needs, or \emph{load-balancing} the tasks on the same set of VMs. E.g., Fig.~\ref{fig:migrate} shows a scaling-in of the $7$ tasks from $5\times 2$-core VMs with $70\%$ utilization to $2\times 4$-core VMs with a $87.5\%$ utilization, a lower billing cost, and also a lower latency due to fewer network hops. Such \textbf{rebalancing needs} are frequent for latency sensitive streaming applications running on elastic Cloud resources that are paid for by the minute. Two key requirements when performing such rebalancing are the \emph{reliability} of messages and task states, and the \emph{rapidity} of completing the migration so that the new deployment stabilizes.


Current DSPS expect the users to decide when to perform such a rebalance, and this typically requires the dataflow to be stopped and restarted with the updated schedule. E.g., Storm's \textbf{\texttt{rebalance}} command allows users to scale-out or -in the number of worker slots assigned to a running dataflow. 
%
However, tasks that are being migrated will loose their state and any messages in their input queue. Users can specify a \emph{timeout} duration, during which Storm pauses the source task(s) so that no new messages are emitted, and in-flight messages may flow through the dataflow.
Users may under- or over-estimate this timeout, causing messages to be lost or the dataflow to be idle, respectively. After the timeout, tasks being migrated are killed and respawned on the new workers, and the source tasks resume generating the input stream. Meanwhile, tasks not being migrated continue to execute while buffering messages in their queues.

There are two capabilities of Storm that can mitigate the impact of lost messages and task states: \emph{message ``acking''} and \emph{checkpointing}. These can ensure reliability but not performance. Storm can guarantee \emph{at least once} message processing using an \textbf{Acknowledgment Service}~\footnote{\href{http://storm.apache.org/releases/1.0.3/Guaranteeing-message-processing.html}{Guaranteeing Message Processing, Apache Storm Version: 1.0.3}}. Each event generated at the source task registers its 64-bit unique event ID with this acking service and this forms the root of a causal tree that is maintained. This event is also temporarily cached at the source. Any downstream events causally generated by tasks processing this root event will add their ID to the tree and acknowledges processing of their parent event by the task. 
The tree itself is compactly maintained by the service using an \texttt{XOR} hash of the each event ID with the root ID, once when the event is added and once when it is acknowledged. Hence, when all causal events are acked, the tree's hash will become zero as each ID is \texttt{XOR}ed twice. Storm checks if the hash for a root event has not become zero within a specified timeout, upon which the event is replayed by the source task. Events whose hashes become zero are periodically discarded from the cache.

The recent Storm versions since \texttt{v1.0.3} support \textbf{checkpointing and recovery} of the state of tasks~\footnote{\href{http://storm.apache.org/releases/1.0.3/State-checkpointing.html}{Storm State Management, Apache Storm, Version: 1.0.3-SNAPSHOT}}. 
Users explicitly implement stateful tasks with interfaces to acquire and restore its state. The framework uses these for periodic distributed checkpointing of tasks, similar to a three-phase commit. A special source task sends a wave of checkpoint messages that flow through the dataflow and triggers these methods in each task, causing state transitions. 
The \texttt{PREPARE} message is sent when a wave is starting, and each task's \emph{prepare} method assembles its internal state. Once all \texttt{PREPARE} messages are acked, the \texttt{COMMIT} message is sent and causes the task's \emph{commit} method to persist the prepared state to an external key-value store (Redis). A \texttt{ROLLBACK} message is sent if the prepare message was not acked for any task. Once committed, the checkpointed states can be restored in future by sending an \texttt{INIT} message. On receiving this, a task's \emph{init} method is passed the last committed state from the external store.

Using these two features, one can perform reliable rebalancing out-of-the-box with in Storm, which we term as \textbf{Default Storm Migration (DSM)}. Here, we can initiate Storm's \texttt{rebalance} command immediately on user request for the new schedule. This will kill all migrating tasks and cause in-flight events to be lost, but the acking service will replay the lost events once the rebalance is completed. Similarly, the checkpointing service will restore the tasks' states from the last periodic checkpoint after the tasks are migrated.

While one is assured of reliability, \emph{this comes at the cost of performance when using DSM}. The number of lost events and state restoration can be disruptive, and delay the dataflow resuming its stable execution. The dataflow snapshot effectively rolls back to the older of the last successfully processed message or the last successful checkpoint. The granularity of event recovery is a root event in the causal tree. So, once the migration is complete, the root events for all in-flight events that were lost will be replayed from the source task and even causal events that were successfully processed earlier will be regenerated and reprocessed. This also means that the old replayed events will be interleaved with the new events being generated by the source tasks immediately after the dataflow is restarted and the source tasks unpaused. 
While Storm and most DSPS do not guarantee event ordering, this interleave will cause a significant number of events to be out of order. 

Both acking and checkpointing need to be on all the time. Acking is done for all events, and the checkpoint interval is periodic ($30~secs$, by default) and has to be configured to balance operational costs and rollback loss for a dataflow. Hence, they also pose additional overheads if the fault-tolerance is a concern only during active migration and not during regular operations~\cite{fischer:bigdata:2015,storm:acking}.  This can be punitive during normal operations if the input rates are high~\cite{chintapalli:ipdps:2016}. One advantage of DSM is that the new schedule is initiated immediately by killing the migrating tasks, with the consequences on recovery pushed to after the rebalance has completed.


\section{Dataflow Migration Strategies}
\label{sec:approach}
In this section, we propose two approaches that address the performance limitations of the baseline DSM strategy. These actively manage the checkpointing, acking and rebalancing to improve the efficiency and timeliness of the dataflow migration, while still guaranteeing reliability. We discuss these conceptually as generic strategies for DSPS, and offer linkages specific to Storm by extending the framework's capabilities.


\subsection{Drain, Checkpoint and Restore (DCR)}
\label{sec:ccr}

\begin{figure}[t]
	\centering
	\includegraphics[width=1.0\columnwidth]{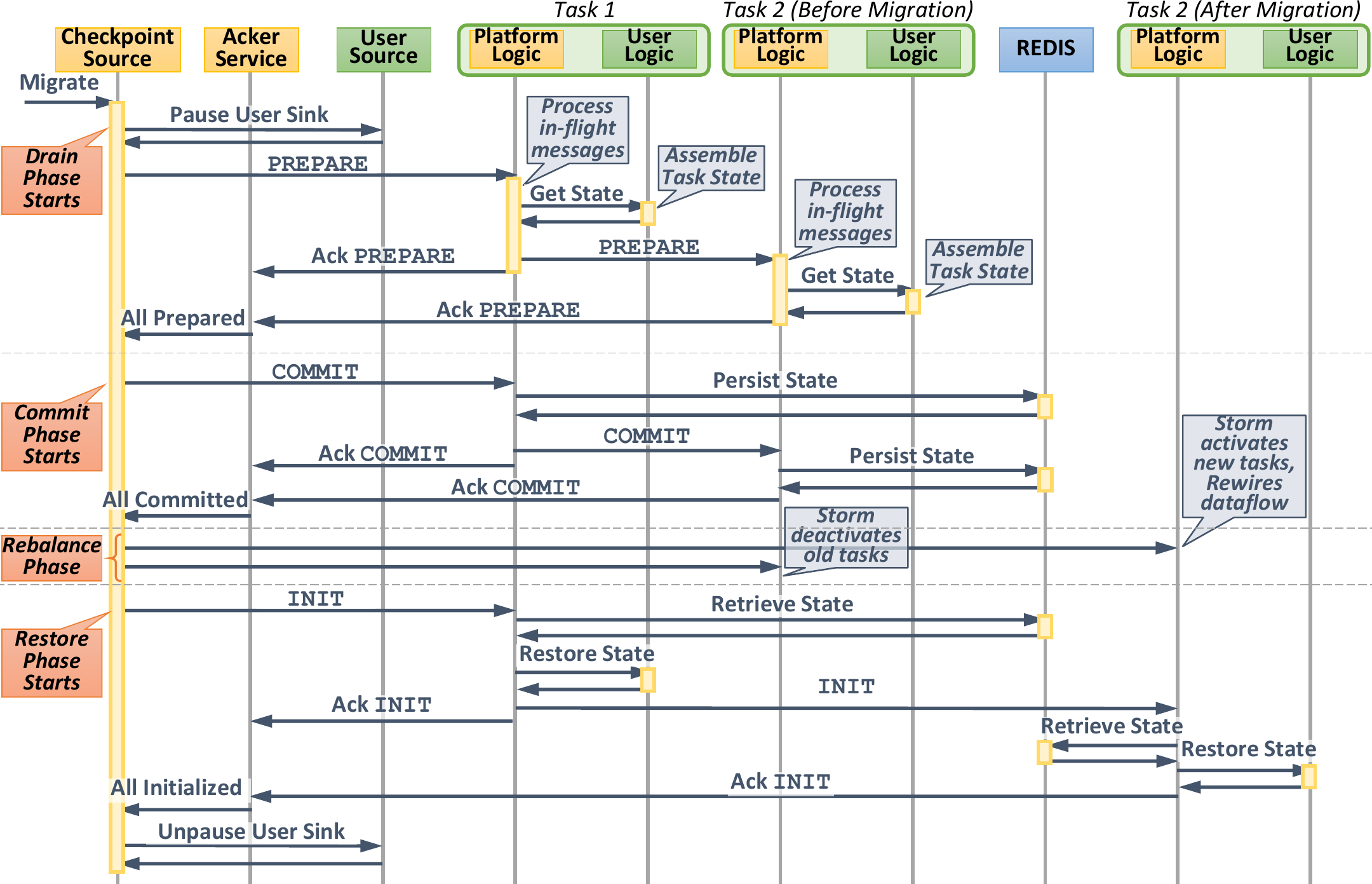}
\caption{Sequence diagram for DCR migration operations}
	\label{fig:dcr:seq}
\vspace{-0.15in}
\end{figure}

Conceptually, our DCR strategy performs three operations to addresses some of the performance limitations of DSM. One key intuition is to pause the source tasks' execution and let the in-flight messages execute to completion across the dataflow. This effectively \emph{drains} the dataflow without any loss of messages before the tasks are migrated, and there are no failed messages to be replayed later. Further, DCR also performs a \emph{just-in-time (JIT) checkpointing} wave after the drain and before the task migrations are done, rather than be enabled periodically. This ensures that the latest state is checkpointed, and we restore the most recent state after the rebalance. Lastly, message reliability is enabled only for the checkpoint messages rather than for all dataflow events. The last two also avoid the overheads for reliability if the user does not require them for normal operations.

\begin{figure*}[t]
	\centering
	\subfloat[DCR]{	
		\includegraphics[width=0.45\textwidth]{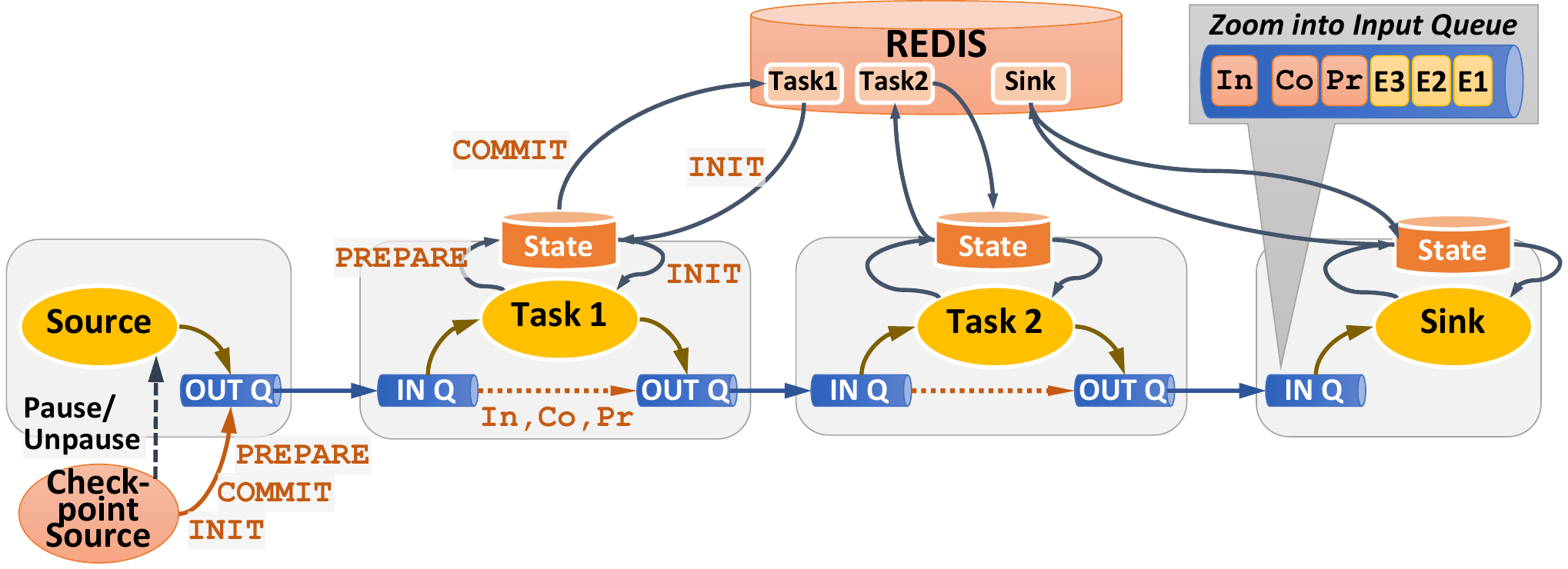}
		\label{fig:dcr:flow}
	}~~
        \subfloat[CCR]{	
		\includegraphics[width=0.52\textwidth]{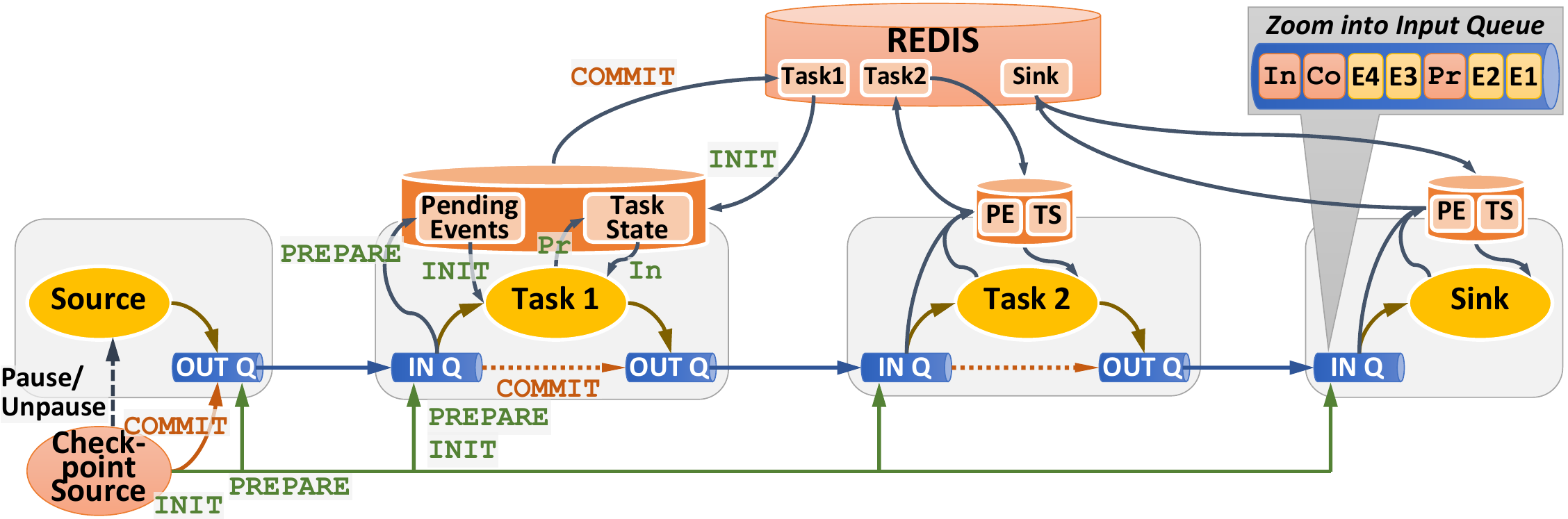}
	\label{fig:ccr:flow}
      }
\caption{Flow of user and checkpoint events, and operations for our two strategies, for a sequential dataflow with 4 tasks.}
\vspace{-0.15in}
\end{figure*}

In the context of Storm, the drain and checkpointing phases are slightly involved and are discussed here. Enabling acking only for reliability of checkpointing messages is simply done by assigning an event ID to the checkpoint events while emitting them for tracking by the acker service. Fig.~\ref{fig:dcr:seq} shows the sequence diagram of operations performed as part of our DCR strategy and Fig.~\ref{fig:dcr:flow} shows the architecture interactions.

When a migration enactment request is received from the user, the schedule planning has already taken place and the new mapping of tasks to VMs decided. We first pause the source tasks from emitting new events. We override the logic for the \emph{Checkpoint Source Task} and initiate a checkpoint wave by sending a \texttt{PREPARE} event. By default, these events flow along the same edges as the original dataflow. Since we have paused the source task, these \texttt{PREPARE} events will be the \emph{last event in the input queue} for every task in the dataflow. The input queue for each Storm worker is single-threaded. So when the task sees the \texttt{PREPARE} event, it knows that it has processed all in-flight data events and the task holds the latest state. Intuitively, the \texttt{PREPARE} event is the rearguard that sweeps behind the data events and guarantees that the dataflow has been drained.

User's task logic extends a Storm platform task class, and this platform logic handles the checkpoint events. When the \texttt{PREPARE} event is seen, the default platform logic calls the user logic to retrieve a snapshot of the current task state. It then forwards the event to its downstream children, and then acks the completion of its processing of the \texttt{PREPARE} event. This happens for every task, and once it reaches the last sink task, all in-flight events have been processed, all task states have been snapshotted, and the acking service notified of this.

After the prepare is completed, our checkpoint task initiates a \texttt{COMMIT} event that flows through the dataflow in a similar manner. The receipt of this message causes the platform logic in each task to persist the task's state snapshot to a Redis distributed store. This event too will be forwarded, and then acked by each task. 
When all \texttt{COMMIT} events have been acked, the checkpoint task invokes the native \texttt{rebalance} command of Storm with a zero timeout. Tasks that need to be migrated are killed and restarted on new slots, and rewired together to form the original dataflow. There will be no messages in-flight to be lost.

Once the rebalance completes, the checkpoint task sends an \texttt{INIT} event through the dataflow to initialize the restarted tasks. This event is again forwarded through the dataflow and serves as the vanguard message in the rebalanced dataflow. When received, the platform logic in each task will retrieve the checkpointed task state from Redis and call the user's logic to restore the state. The \texttt{INIT} events are acked after forwarding as well. It may so happen that the task is not ready when an \texttt{INIT} event is forwarded to it within the $30~secs$ default acking timeout.
To avoid rolling back the rebalance when \texttt{INIT} events are not acked, the checkpoint task emits duplicate \texttt{INIT} events after each $1~sec$, and the platform logic at a task skips processing this event if the task has already restored its state. Once all tasks have acked an \texttt{INIT} event, the source task is unpaused and resumes generating messages through the rescheduled dataflow.

DCR has several advantages over DSM. It avoids event loss by draining the dataflow completely before the rebalance. As a result, there is no need to replay lost messages, which avoids their reprocessing costs. There is also a clear boundary between events processed by the dataflow before and after the migration, 
with no interleaving of old and new messages. We also avoid costs of periodic checkpointing by doing a JIT wave. 
One downside of the DCR approach is the time spent in draining the dataflow, during which only some of the tasks are processing events while the rest are idle. 
This depends on the number of in-flight messages, and the checkpoint can start only after these prior events are processed. Further, input events that are queue up within the source tasks have to flow through the rebalanced dataflow to catchup. We address some of these in the next proposed strategy, CCR.


\subsection{Capture, Checkpoint and Resume (CCR)}
\label{reb:ccr}

We address the key limitation of the DCR approach, which is the time spent in draining the dataflow of all in-flight messages before the checkpoint starts. There are two aspects to this drain time. (1) The checkpoint messages flow incrementally through the dataflow to guarantee that they are the last event, and hence take additional time to reach the sink tasks from the source. (2) All in-flight messages have to be fully processed by all dataflow tasks before the rebalance can begin. Both of these incur additional latency.




We address the first challenge by directly \emph{broadcasting} checkpoint events from the source task to each task in the dataflow.  
This hub-and-spoke model allows the checkpoint events, specifically \texttt{PREPARE} and \texttt{INIT}, to be directly placed at the end of each task's input queue. As a result, we avoid having to pass the event through all the preceding tasks and their processing logic. We retain the sequential wiring for \texttt{COMMIT} events to ensure that all in-flight user events in the input queues have been handled.

We address the second problem by processing only the one possible event that a task is currently executing and \emph{capturing} all other messages that are in its input queue, for every task in the dataflow. We also do not emit any events downstream after processing, and instead capture the output events as well. This effectively takes a snapshot of all in-flight messages and pauses further flow through the dataflow. The most time that this will take is the time taken by the slowest task to drain is local input queue until the broadcasted \texttt{PREPARE} message is seen. In contrast, DCR takes time for every in-flight event across all task queues to be processed by every downstream task in the dataflow. This snapshot of the input and output events by CCR for each task is appended to the state of the task, and restored into the input and output queues after the dataflow is rebalanced. This allows the dataflow to \emph{resume} execution of the in-flight events.

These have to be carefully coordinated to guarantee consistency and reliability. We discuss these here, along with its implementation within Storm, and this is illustrated in Fig.~\ref{fig:ccr:flow}. Besides the checkpoint source task, we also extend the base platform logic for each task for the CCR strategy.

The series of steps are similar to DCR. When the dataflow starts, we wire the checkpoint source task to every task in the dataflow as a broadcast channel. When the user's migration request is received, we pause the source task(s) and send a \texttt{PREPARE} event on the broadcast channel to every task. This event is received and placed at the end of the input queue. Each task continues to process user events received ahead of the \texttt{PREPARE} in their input queue, and emit output events as well after processing. Hence, the \texttt{PREPARE} event may be at any position within the input queue.

When the \texttt{PREPARE} event is processed from the top of the queue by our platform logic in the task, we enable a capture flag. This flag ensures that future events seen on the input queue are added to a \emph{pending event list} without being processed, and this list is appended to the task's state. The \texttt{PREPARE} event is then acked, but not forwarded downstream.

When the checkpoint task receives acks from all tasks for the \texttt{PREPARE} event, it sends a \texttt{COMMIT} event, but as part of the dataflow's wiring. This causes the \texttt{COMMIT} to sweep through the dataflow and is guaranteed to be the last event in the input queue for every task. On receipt of this event, our platform logic at a task persists the task's user logic state as well as the pending event list to Redis. The \texttt{COMMIT} is forwarded downstream, and then acked. Once all tasks have acked the \texttt{COMMIT} event, we have successfully captured all in-flight messages and the user's task state.

As for DCR, we then initiate Storm's \texttt{rebalance} command with zero timeout, and once that is done, broadcast an \texttt{INIT} event from the checkpoint task to all tasks in the rebalanced dataflow. This will be the first event in the input queue for tasks in the dataflow. Now, the goal is not just to restore the tasks' user logic state but also for the captured events to resume execution. When \texttt{INIT} event is seen, our platform logic at each task fetches the task's state from Redis. The user state is passed to the initializer of the user task. The pending event list is then replayed locally to the task logic for processing and the generated output events are sent to downstream tasks. The \texttt{INIT} event is acked by the task and then all events in the pending list are processed.
When all acks for the \texttt{INIT} event is received, we unpause the source task(s) in the dataflow to resume generation of new events.

As can be seen, this CCR approach addresses the shortcomings of the DCR strategy while retaining several of its advantages. We reduce the drain time of DCR significantly, allowing the rebalance to be enacted more quickly and with fewer messages to queue up at the source tasks. Intuitively, CCR overlaps the dataflow drain time of DCR with the dataflow refill time after rebalance to offer benefits. But it does not eliminate the drain time completely like DSM. We do pay additional overhead to send the state of the captured events to Redis and to restore them, but this is still cheaper than replaying/reprocessing them and their ancestors in the causal tree, like DSM does. 

\begin{figure*}[t!]
	\centering	
        \includegraphics[width=0.95\textwidth]{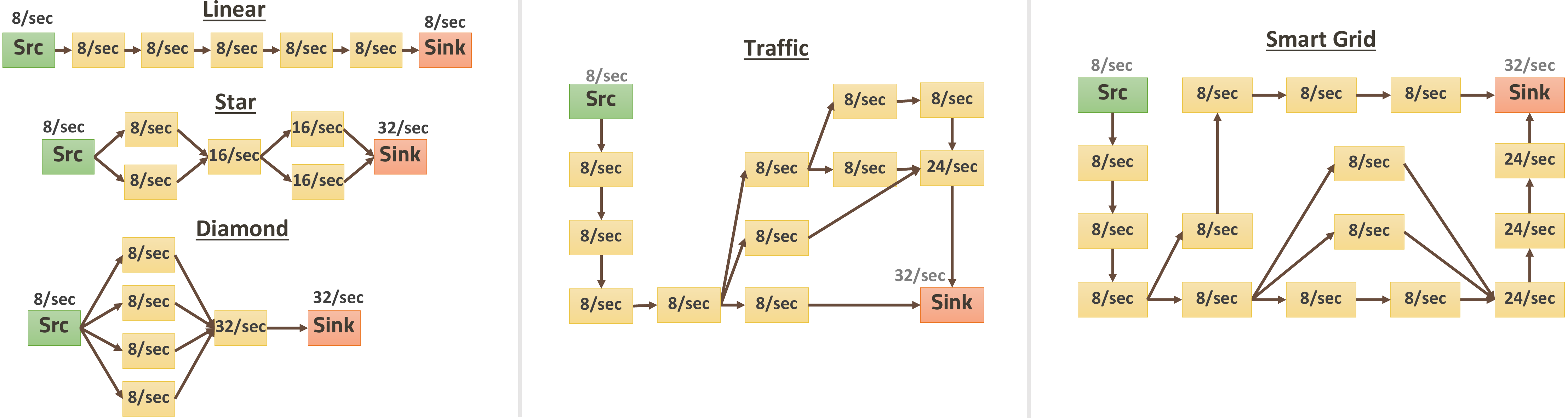}
	\caption{Micro and Application DAGs used in experiments. Cumulative input to each task is indicated within each.}
			\label{fig:dag}
\vspace{-0.15in}
\end{figure*}

\section{Performance Metrics}
\label{sec:metrics}

We propose several performance metrics that should be considered when evaluating the effectiveness of these strategies. Since all these approaches guarantee reliability and consistency, that is not listed as a separate metric.

\begin{enumerate}[leftmargin=0.5cm,itemindent=-0.3cm,labelsep=0.2cm,align=left]
\defn{Restore Duration}{This is the time taken from the start of the user-initiated migration request, to the first message being seen in any of the sink tasks. During this period, there will be no output events that come out of the dataflow (i.e., output throughput is 0). This is applicable to all strategies.}

\defn{Drain/Capture Duration}{This is the time duration for the DCR and CCR strategies when the dataflow is being drained, and the task and message states are being persisted, after the migration request is received from the user. After this duration, Storm's \texttt{rebalance} command is initiated. This time is not applicable (i.e., $0$) for DSM.}

\defn{Rebalance Duration}{This is the time taken for Storm's \texttt{rebalance} command to complete. This initiates the kill of the tasks being migrated, and their redeployment on new machines. When completed, tasks of the dataflow are being started on the new VMs and waiting to be initialized with \texttt{INIT} events.}

\defn{Catchup time}{This is the time point when all old messages that had entered the dataflow before the migration was initiated have been successfully processed and emitted from the sink of the dataflow after its migration. This is relevant only for DSM and CCR, and not for DCR since it drains all old events before the migration.}

\defn{Recovery time}{This is the time point when all new messages that enter the dataflow after migration \emph{and failed due to timeouts} have been successfully processed and emitted from the sink of the dataflow. After this point, we do not see any loss/recovery of new messages due to the migration. This is only applicable for DSM as there will not be any failed messages in DCR and CCR.}

\defn{Rate Stabilization time}{This is the time point after which the output message rate of the dataflow remains stable after migration, and consistent with the expected stable input rate. We define stability as being achieved when the observed output rate is sustained within 20\% of the expected output rate for $60~secs$. The start of this stable time window indicates stabilization.}

\defn{Message Loss/Recovery Count}{This is the number of messages that were lost \emph{and recovered} after migration due to killing the dataflow or acking timeouts.}
\end{enumerate}




\section{Results}
\label{sec:results}


We validate our proposed migration approaches, DCR and CCR, on the Apache Storm DSPS, and compare it with the default Storm migration (DSM). 

\textbf{Implementation.} We implement the migration strategies on Storm $v1.0.3$. For DCR and CCR, our custom checkpoint source task is implemented by overriding the \texttt{CheckpointSpout} class. 
For CCR, we further modify the \texttt{TopologyBuilder} class to automatically create the broadcast wiring from the checkpoint source to all other tasks. 
We also modify the \texttt{StatefulBoltExecutor} class that each user task extends, to support the capture and resume capabilities of CCR. For DSM, we enable periodic checkpointing with the default interval of $30~secs$, and enable acking for all events. We use Storm's \texttt{rebalance} command with a timeout of $0$ in all cases. Redis $v3.2.8$ is used to persist the checkpoints using native Storm bindings. 

\textbf{System Setup.} A Storm Cluster is deployed on Microsoft Azure D-series VMs in the Southeast Asia data center. The type and number of VMs vary with the experiment (Table~\ref{tab:vmcount}) and range from $2-21$~VMs with $1-4$~cores each. Each resource slot of Storm runs a distinct task instance, and is assigned a 1-core \emph{Intel Xeon E5 v3 CPU @2.4 GHz} with \emph{3.5GB RAM}. A \emph{50~GB SSD} and \emph{1~Gbps Ethernet} is shared by all slots. Redis runs on a seperate D3 VM with $4~cores$. 

\textbf{Application Setup.} We use two types of streaming dataflows in our experiments, \emph{micro-DAGs} and \emph{application DAGs}, as illustrated in Fig~\ref{fig:dag}. The micro-DAGs capture common dataflow patterns seen in Streaming applications, and are often used in literature~\cite{xu:ic2e:2016,millwheel:vldb:2013,peng:middleware:2015}. \emph{Linear}, \emph{Diamond} and \emph{Star} respectively capture a sequential flow, a fan-out/-in, and a hub-and-spoke pattern, with $5$ user tasks each, besides a source and a sink.
We use two application DAGs with structures based on real-world streaming applications. \emph{Traffic}~\cite{biem:sigmod:2010} analyzes the traffic patterns from GPS sensor streams, and \emph{Grid}~\cite{simmhan:cise:2013} does predictive analytics over electricity meter and weather event streams from Smart Power Grids.
For simplicity and reproducibility, we use a dummy task logic with a sleep time of $100~millisecs$ for all the tasks since it is orthogonal to the behavior of the strategies. 

All tasks have a \emph{selectivity} of 1:1, i.e., one output event is generated for one input event. In our experiments, the source task generates synthetic events at a fixed $8~events/sec$ rate, which is $20\%$ less than the $10~events/sec$ peak supported rate per task instance, given a $100~ms$ task latency. The input rate at each task goes up to $32~events/sec$ for our DAGs (Fig~\ref{fig:dag}).  
We assign one task instance (thread) for each incremental $8~events/sec$ input rate to a task, with each task instance allocated one exclusive resource slot for execution. 


\begin{table}[b]
	\footnotesize
	\centering
	\caption{Tasks, slots and VMs for the Dataflows}
	\label{tab:vmcount}
	\begin{tabular}{l||l|p{.95cm}|p{1cm}|p{1cm}|p{1.1cm}}\hline
		\textbf{DAG}          & Tasks$^*$ & Task Instances (Slots) & \emph{Default} \#VM w/ $2$ slots & \emph{Scale-in} \#VM w/ $4$ slots &\emph{Scale-out} \#VM w/ $1$ slot \\ \hline  \hline                                              
		\textbf{Linear}     & 5         & 5     & 3        & 2 &  5         \\\hline 
		\textbf{Diamond}    & 5         & 8      &4        & 2  &  8   \\\hline 
		\textbf{Star}  & 5         & 8      & 4        & 2      &  8   \\\hline 
		\textbf{Grid}  & 15         & 21      & 11        & 6& 21       \\\hline 
		\textbf{Traffic}  & 11         & 13      & 7        & 4   &   13    \\\hline 
		 \hline                                   
		 \multicolumn{6}{l}{\emph{$^*$ Excludes Source and Sink tasks that are on a separate $4~core$ VM}}
	\end{tabular}
\end{table}

\textbf{Experiment Setup.} We evaluate the migration strategies for the two most common elasticity scenarios in Clouds: \emph{scale-in} and \emph{scale-out} of the number of VMs. 
For scale-in, we initially deploy the dataflows on $n$ D2 VMs with $2$ slots each, and then migrate the dataflow to $\lceil \frac{n}{2} \rceil$ D3 VMs with $4$ slots each. In the scale-out experiments, we go from $\lceil \frac{n}{2} \rceil$ D2 VMs with $2$ slots to $n$ D1 VMs with $1$ slot each. The total number of slots used does not change, just the VMs they are packed on. 
The source and sink are both assigned to a single 4-slot VM. They are not migrated, to allow logging of end-to-end statistics without time-skews. Storm's default round-robin scheduler is used to map a task instance to an available VM slot, during initial deployment and on rebalance.

Each experiment is run for $12~mins$ and the user migration is initiated $3~mins$ after the dataflow submission to ensure a stable start.  We log the timestamp of checkpoint and user events from Storm and on the VMs to help evaluate the metrics that we have proposed.

\subsection{Analysis}

\begin{figure}[t!]
	\centering
\subfloat[Scale-in]{
	\includegraphics[width=0.45\textwidth]{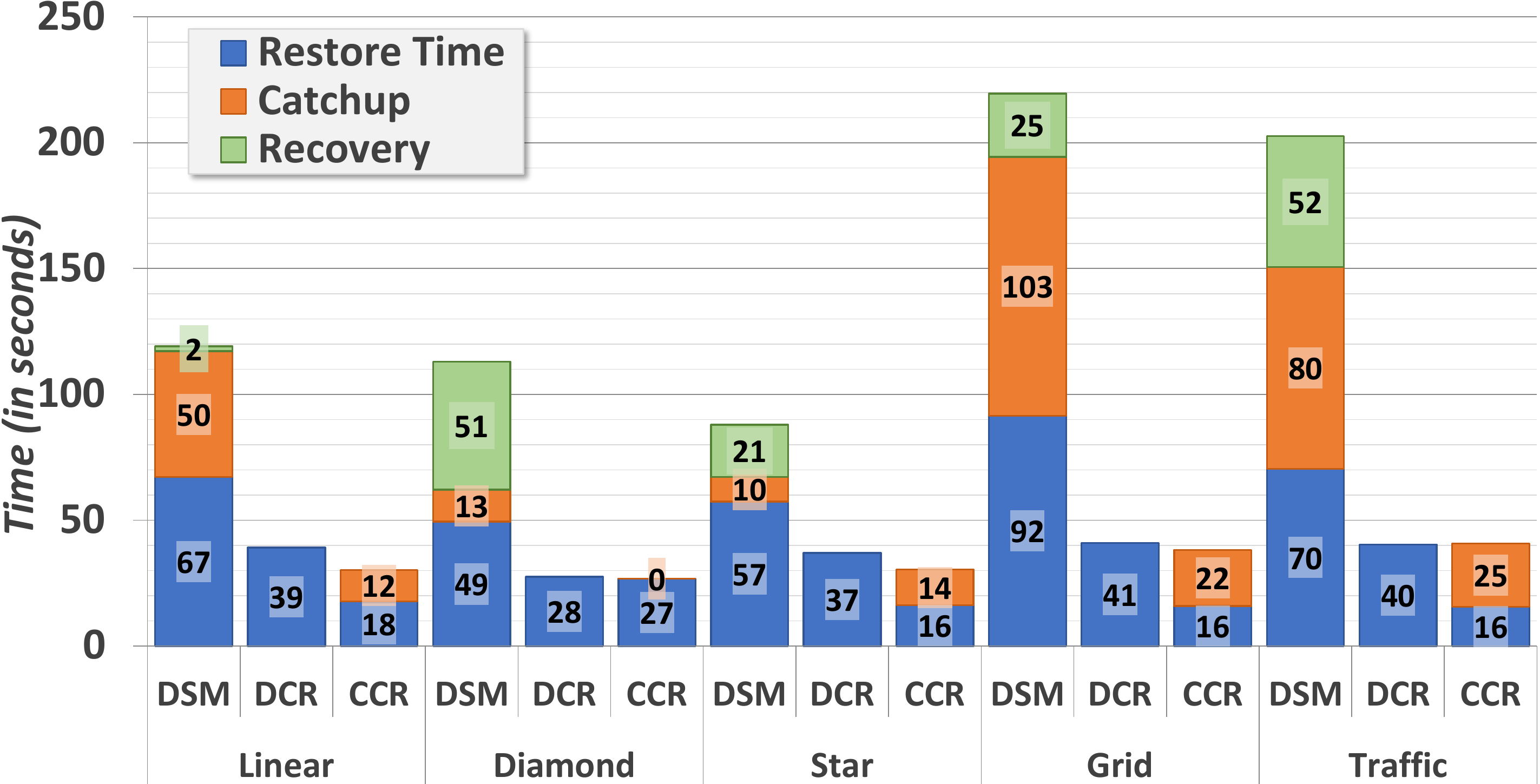}
	\label{fig:up:timing}
      }\vspace{-0.1in}\\
      \subfloat[Scale-out]{
		\includegraphics[width=0.45\textwidth]{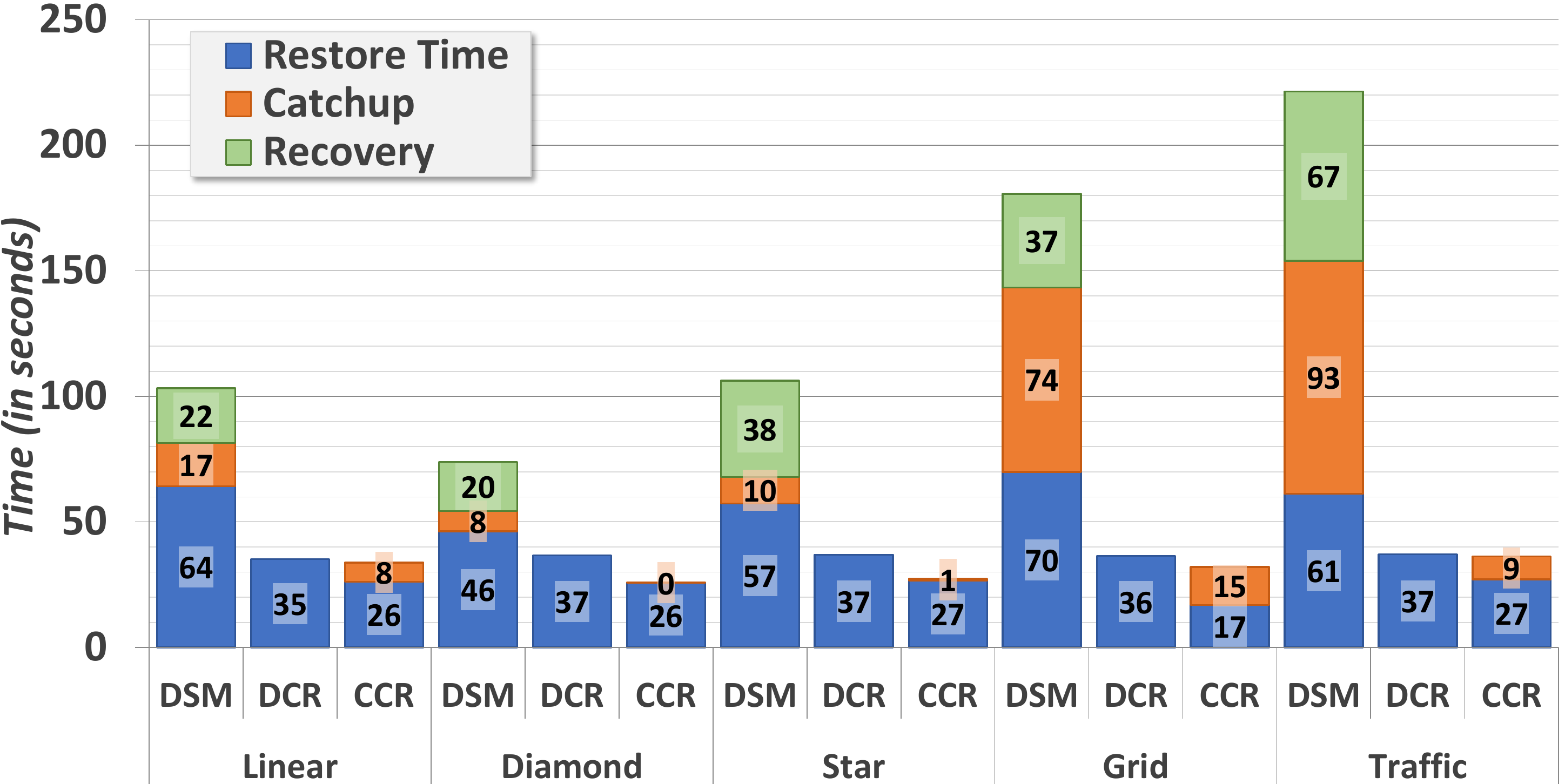}
		\label{fig:down:timing}}\\

	\caption{\emph{Performance time} for different strategies}
		\label{fig:timing}
\end{figure}

We evaluate and compare the three migration strategies, DSM, DCR and CCR, based on the proposed metrics, for the $5$ dataflows and two scaling scenarios outlined above.


Fig.~\ref{fig:timing} shows a stacked bar plot with the time taken (Y axis) for restore, catchup and recovery for the $3$ strategies and $5$ DAGs. 
These three are user-facing quality metrics that are common to all approaches, and are visible sequentially from the migration request time.
%
We see that \textbf{Restore time} is consistently the least for CCR, followed by DCR and DSM. This holds across scale-in or out and for all dataflows. E.g., for scale-in experiment for Grid, CCR takes $15~sec$, DCR $41~sec$, and DSM $91~sec$. 

DSM has no drain time. The cause for this delay is due to the \texttt{INIT} events that need to be sequentially sent to each task after rebalance. Further, we observe several cases where the initial \texttt{INIT} events timeout without acking due to the tasks not being active yet, and are resent after a $30~sec$ timeout. This is seen in the restore time growing in $\approx30~sec$ jumps, with each new wave of timed-out \texttt{INIT}s required. While DCR also sends the \texttt{INIT} events sequentially, we aggressively resend them every $1~sec$. While this causes duplicate events (that are ignored if a task is already initialized), these are few enough to justify the benefits of lower initialization delay. As a result, it is able to restore all tasks more quickly than DSM, despite spending additional time draining the dataflow. CCR broadcasts the \texttt{INIT} events, with the same $1~sec$ resend logic, and initiates the execution of the tasks even more quickly. E.g. in the scale-out of the Grid dataflow, the first \texttt{INIT} after the user initiated migration is received by a task at $31~sec$ using DCR, and at $17~sec$ for CCR. The difference is due to the drain time. 
Both DSM and DCR also have overheads for filling the dataflow after the initialization while CCR resumes from prior in-flight state.

The total migration time for DSM in Fig~\ref{fig:timing} is much higher for application DAGs than for micro DAGs. E.g., the scale-in of the Linear DAG requires $\approx{120}~sec$ but Grid requires $\approx{220}~sec$. This is because the Recovery time is higher due to more tasks utilization by \texttt{INIT} events, leading to subsequent impact on the Catchup and Recovery also. Our scaling experiments show that DSM's performance deteriorates with the size of the DAG. However, our DCR and CCR migration approaches are \emph{less sensitive to the size and complexity of the dataflow}, and are able to migrate all DAGs within $\approx{50}~ms$ time.


One of the factors in the restore time is the \textbf{Drain Time} for DCR and CCR. This value is larger for DCR than CCR since the former waits for all events to flow through the DAG and execute while the latter captures events that arrive at a task after the \texttt{PREPARE} event. 
This difference is proportional to the critical path length or latency of the dataflow and the input event rate. 
E.g., scale-in of Grid shows a drain time of $1,875~ms$ for DCR and $468~ms$ for CCR, while it's scale-out drains in $1,440~ms$ and $550~ms$, respectively. However, this delta is smaller for micro-DAGs, with the scale-in of Linear having drain times of $905~ms$ for DCR and $256~ms$ for CCR.
To verify this, we have run experiment for a linear DAG with $50~tasks$. We find that difference in drain times is $4,352~ms$, which is much higher. 
CCR however has to checkpoint the in-flight messages, besides the task state,but this incremental time is small. E.g., micro-benchmarks show that it takes just $100~ms$ to checkpoint $2000$ events to Redis from Storm.

Yet another component of the Restore time is the rebalance duration, when the actual Storm command runs. This time remains relatively constant across dataflows, VM counts and strategies, with an average 
value of ${7.26~secs}$.

\begin{figure}[t!]
	\centering
	\subfloat[Scale-in]{
		\includegraphics[width=0.2\textwidth]{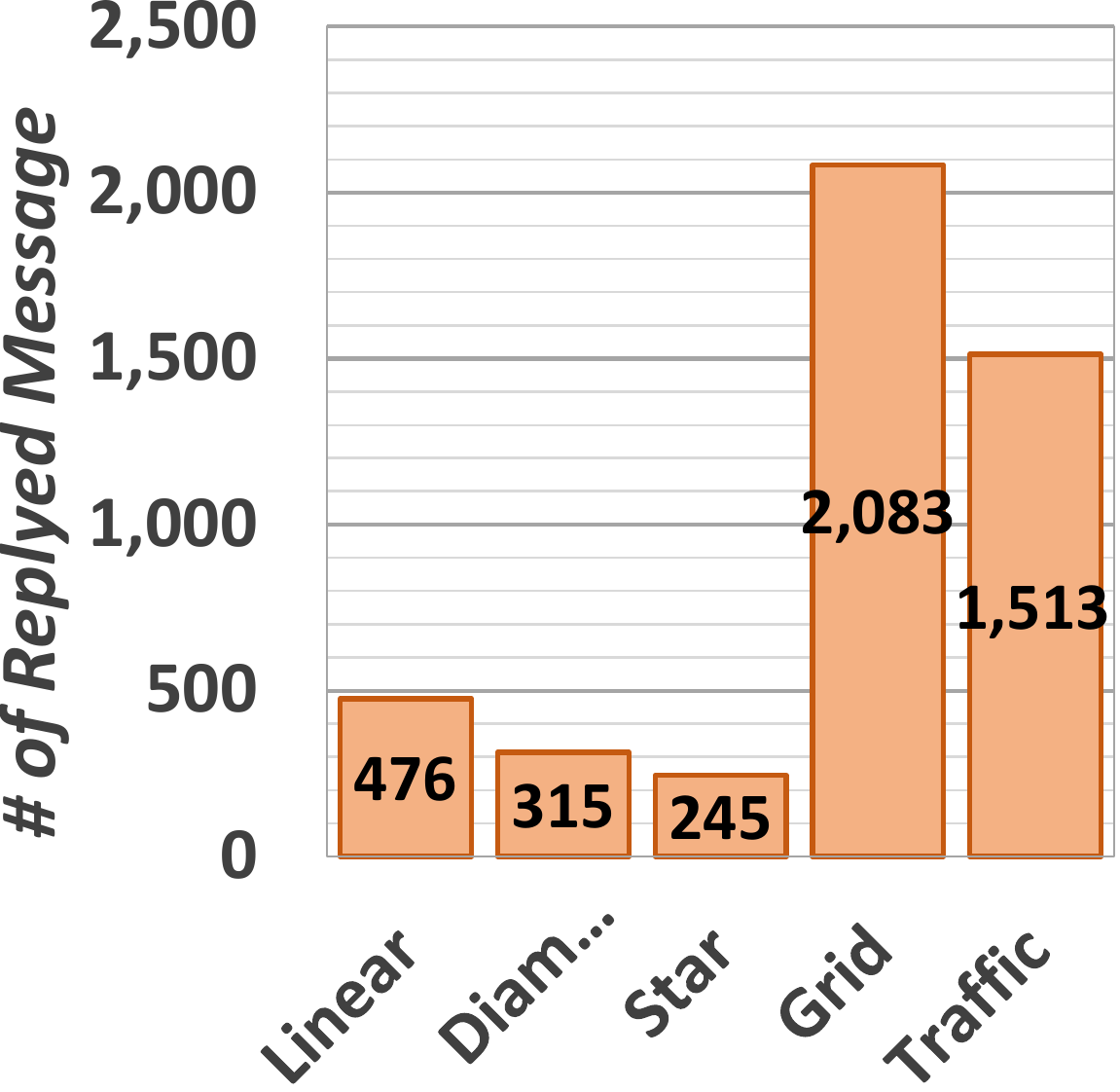}
				\label{fig:up:replay}
	}~~~~
	\subfloat[Scale-out]{
		\includegraphics[width=0.2\textwidth]{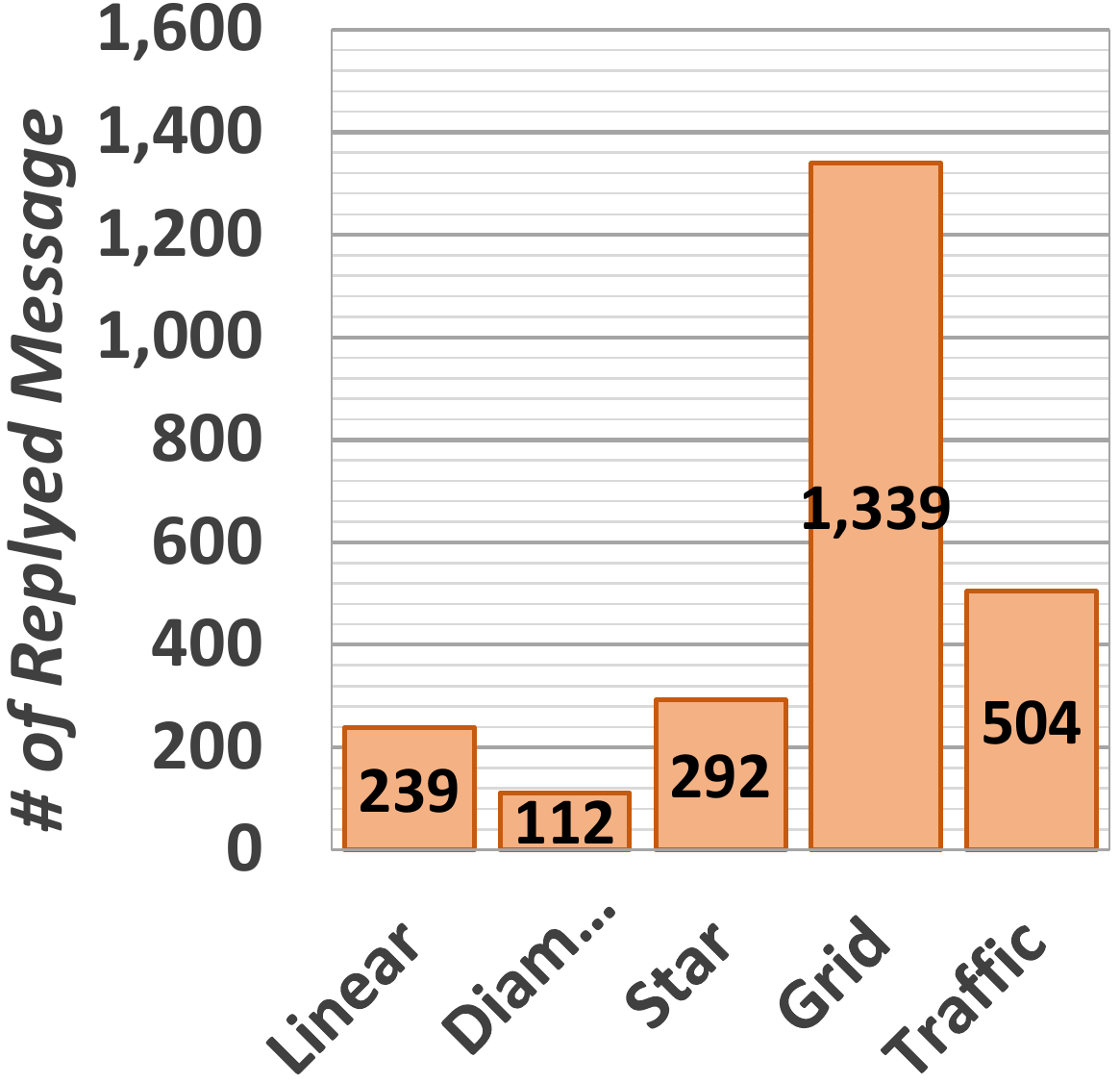}
				\label{fig:down:replay}
	}
	\caption{Number of failed and \emph{replayed} messages for DSM}
		\label{fig:replay}
\vspace{-0.15in}
\end{figure}

\begin{figure}[t]	
	\subfloat[DSM]{
		\includegraphics[width=0.45\textwidth]{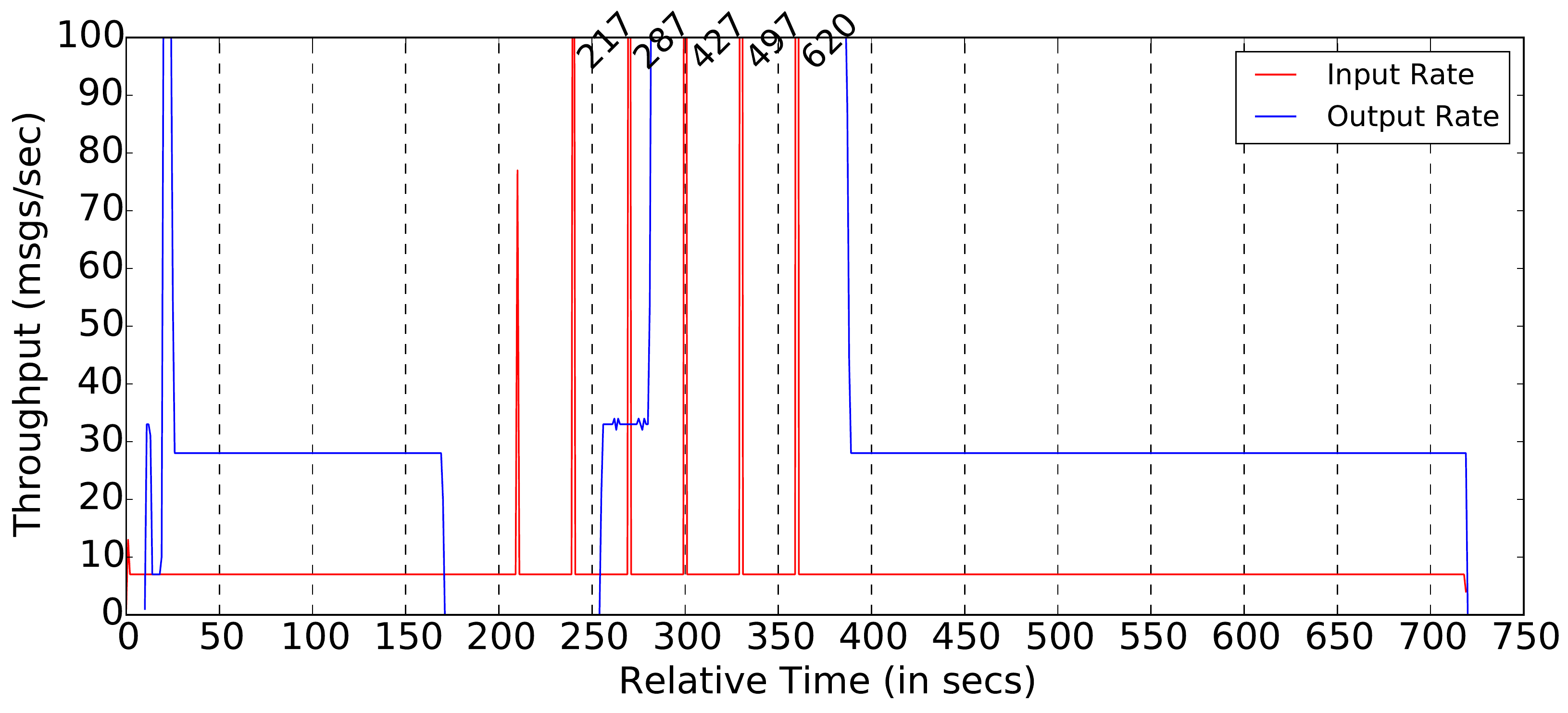}
				\label{fig:up:thru:dsm}
	}\vspace{-0.15in}\\
	\subfloat[DCR]{
		\includegraphics[width=0.45\textwidth]{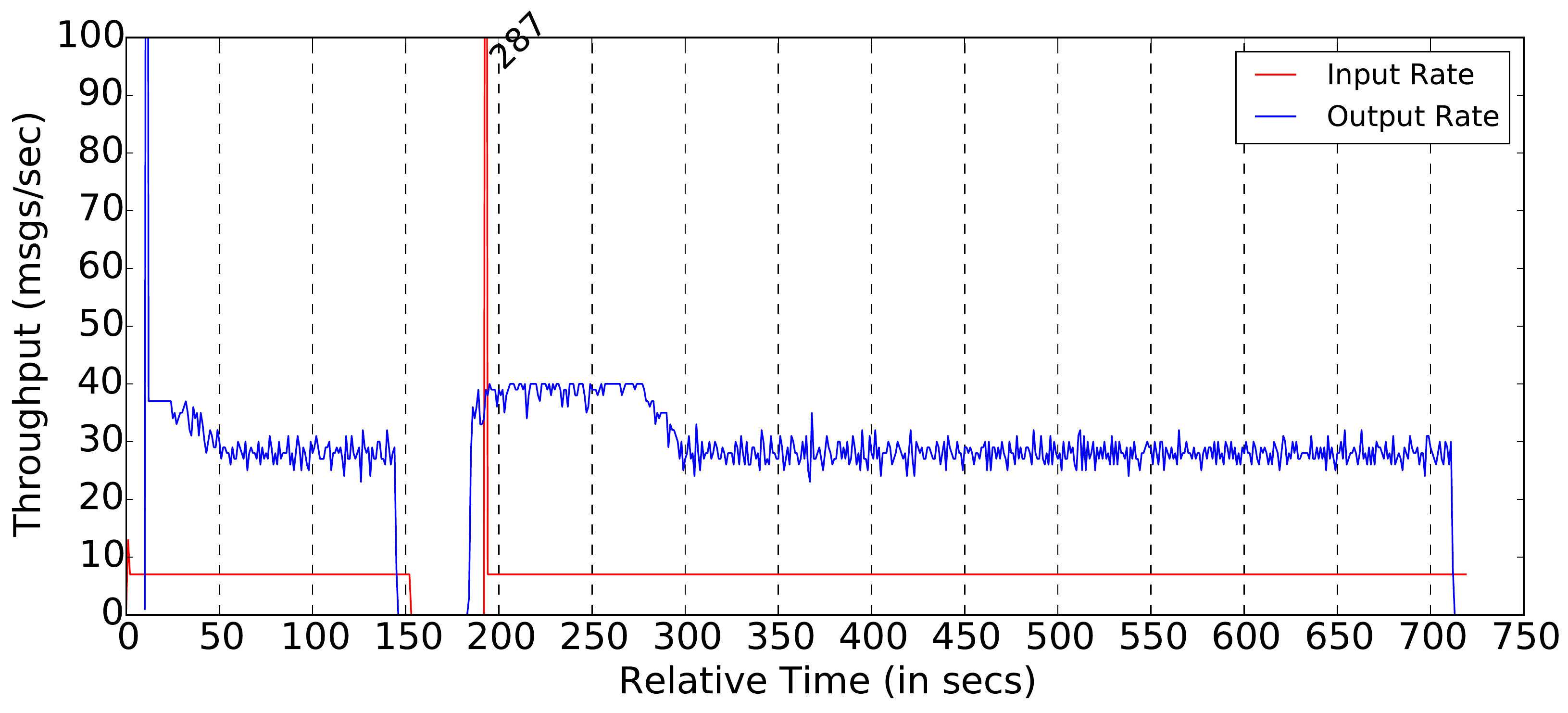}
				\label{fig:up:thru:dcr}
	}\vspace{-0.15in}\\
	\subfloat[CCR]{
		\includegraphics[width=0.45\textwidth]{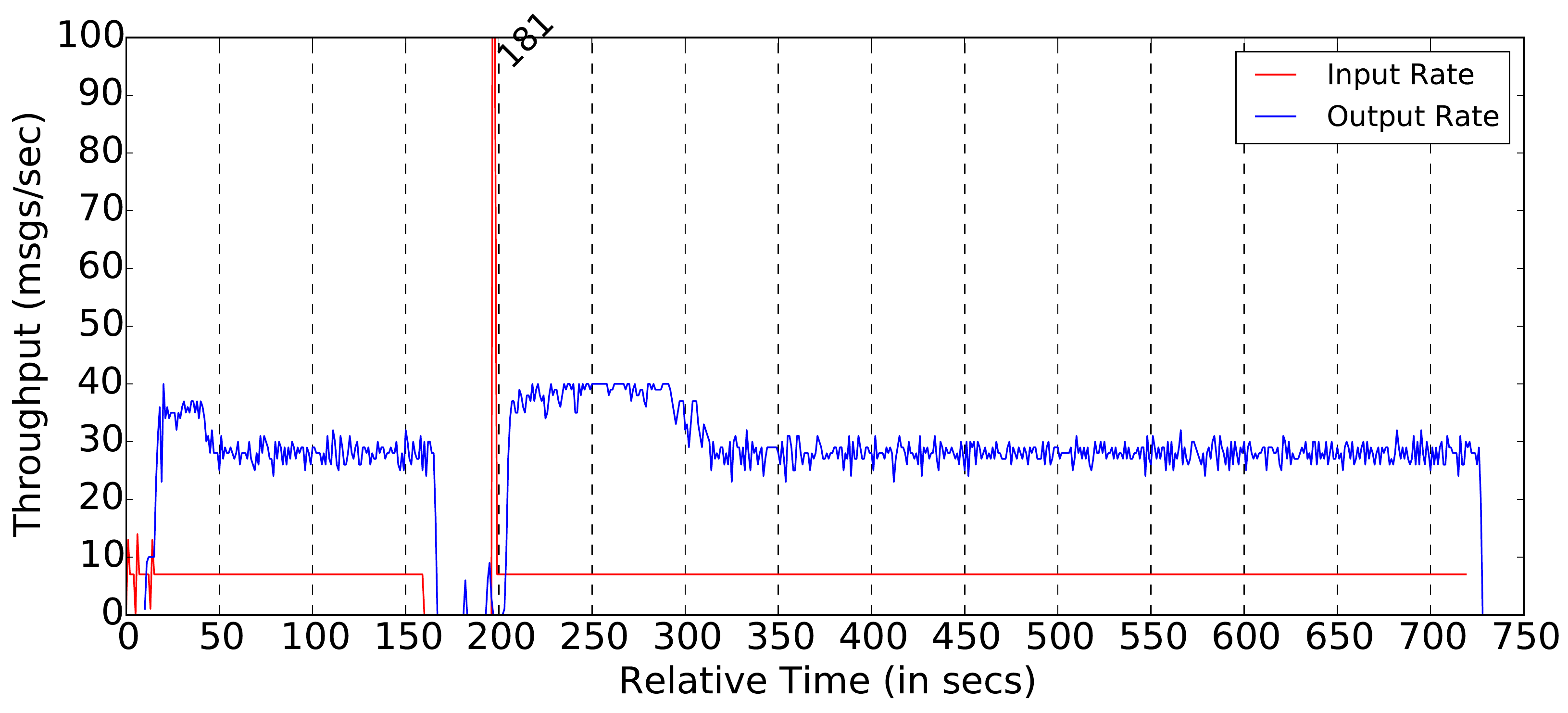}
				\label{fig:up:thru:ccr}
	}
	\caption{\emph{Timeline plot} showing \emph{Input} and \emph{Output} throughput during the \emph{scale-in} of Grid dataflow. Migration request time is shown as ``0'' on X axis.
}
		\label{fig:up:thru}
\vspace{-0.15in}
\end{figure}

The \textbf{Catchup time} in Fig.~\ref{fig:timing} is the time to receive the last old tuple at the sink after migration. This is larger for DSM than CCR, and it is absent for DCR since there are no in-flight events. In DSM, lost events are replayed after their acking timeout. The old events that were discarded due to rebalance will be re-emitted by the source after the $30~sec$ acking timeout occurs. Hence, there is up to $30~secs$ delay for the old events to pass through the dataflow. This is clear when we examine the timeline plot for the input and output event rates shown for the scale-in of Grid dataflow in Fig.~\ref{fig:up:thru:dsm}. We see spikes of the input rate at $30~sec$ intervals after the $200~sec$ point since the migration was requested. The first spike indicates the replay of the old in-flight events. The other spikes are due to the replayed old events or the newly emitted events not being processed quickly by the dataflow due to a high load, and being replayed yet again. The high output rate during this period shows that the dataflow is pushed to its limits. Fig.~\ref{fig:replay} plots the number of such events that are replayed for DSM. These range from $112$ for the scale-out of Diamond to $2,083$ for the sclae-in of Grid. The values are larger for the application DAGs than the micro DAGs more events are timed-out in the larger DAGs, and replayed. 

%
In CCR, catchup time is comparatively smaller as the old events are immediately replayed after being restored from Redis on receiving the \texttt{INIT} event. 
%
The catchup time is higher for application DAGs than micro DAGs. For larger DAGs, the number of in-flight events that are lost are likely to be higher due to more number of tasks/input buffers. Hence, the replayed event count will be more, and they will also require more time to reach to sink from the source task.
%
The catchup time is almost the same for both scale in and out. One may expect some benefits with fewer VMs in scale in due to collocation of tasks that avoids network latency, but the round-robin Storm scheduler may not exploit this. 

\begin{figure}[t!]
	\centering
	\subfloat[Scale-in]{	
		\includegraphics[width=0.4\textwidth]{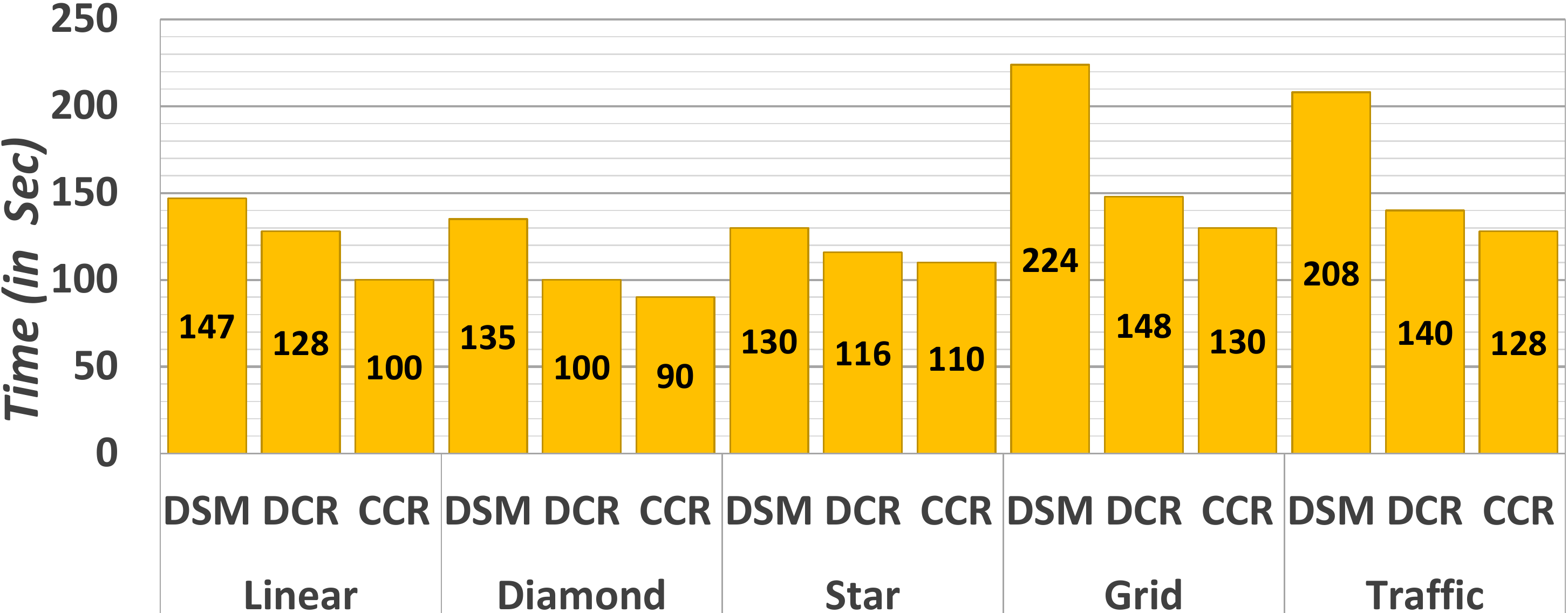}
				\label{fig:up:stab}
	}\vspace{-0.1in}\\	
	\subfloat[Scale-out]{	
		\includegraphics[width=0.4\textwidth]{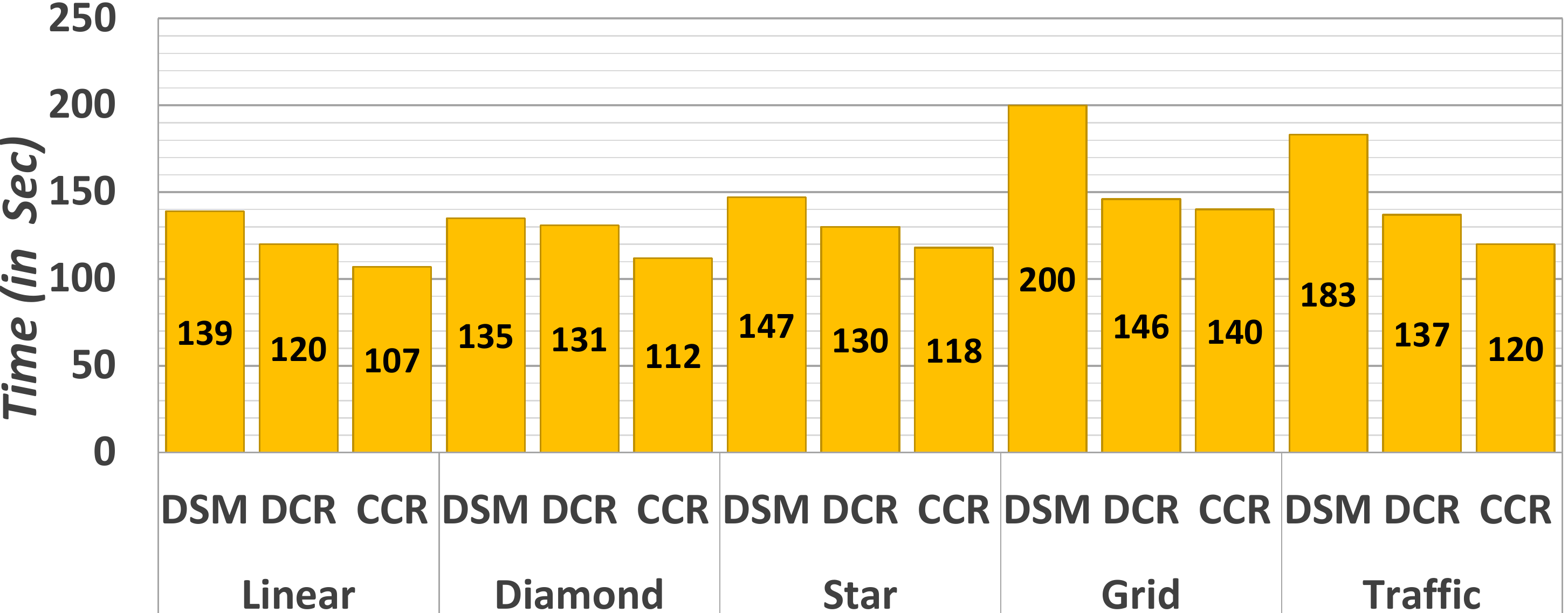}
				\label{fig:down:stab}
	}	
	\caption{\emph{Stabilization time} for different strategies}
		\label{fig:stab}
\vspace{-0.15in}
\end{figure}

Fig.~\ref{fig:timing} shows the \textbf{Recovery time} to receive the last failed and replayed event at the sink, be they old or new replayed events. This is indicative of the dataflow approaching a steady state, and not losing and replaying events due to the migration. 
%
The Restore time is present and high for DSM for all DAGs and for scale-in and out. There is no recovery time required for CCR and DCR since we see no event losses to be replayed. This DSM behavior is a cascading effect of its restore and recovery times, that cause the source task to buffer more events. These events when released overwhelm the dataflow causing event timeouts and replay. 
In fact, the recovery time can be directly co-related with the high \textbf{stabilization time} for DSM in Fig~\ref{fig:stab}, relative to DCR and CCR. In fact, some experiments show DSM take $60~secs$ longer than DCR and CCR to reach a stacbe output rate. 

\begin{figure}[t]	
		\includegraphics[width=1.0\columnwidth]{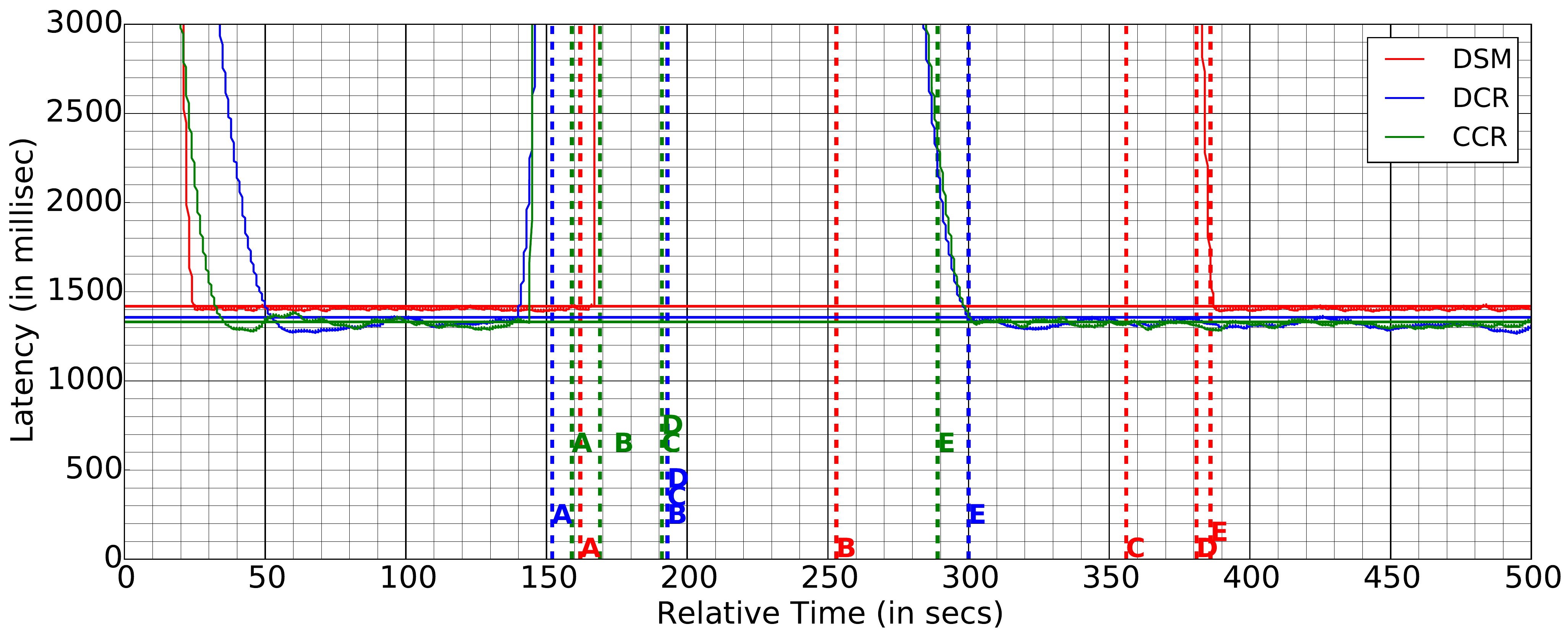}
\caption{Timeline plot showing \emph{Average Latency} over a moving window of $10~secs$ ($\approx{80~events}$) for \emph{scale-in} of Grid dataflow. 
Labels $A..F$ on vertical dashed lines denote \emph{Restore Duration} ($A\to B$), \emph{Catchup Duration} ($B\to C$), \emph{Recovery Duration} ($C\to D$), \emph{Stabilization Time} ($D \to E$).
Horizontal solid RGB lines indicate the stable latency.} 
\label{fig:up:latency}
\vspace{-0.15in}
\end{figure}

Lastly, we analyze the input and output throughputs, and end-to-end latency for the dataflows during migration. Fig~\ref{fig:up:thru} is a timeline plot of the input rate for the dataflow at the source task and the output rate seen at the sink task, during scale-in of the Grid DAG. Time $0$ on the X Axis indicates the start of the migration request. Fig~\ref{fig:up:latency} similarly shows the corresponding average event latency over a window of $10~secs$ seen for output events from the Grid dataflow. 

The throughput plot, Fig~\ref{fig:up:thru}, shows that the steady input rate is $8~events/sec$ and output rate is $32~events/sec$, as the selectivity of the Grid DAG is $1:4$. We observe that the sink task is paused in DCR and CCR during migration, with zero input rate, but not in DSM. This reduces the interleaving of old and new events after migration and avoids event losses due to time-outs. The single input rate peak for DCR and CCR show the backlogged events emitted when sink is resumed. As mentioned before, multiple such peaks exist for DSM. 
The output throughput has a small increase during $\approx 180-300~secs$,  
showing the \texttt{INIT} events flowing, the captured events in Redis replayed for DCR, and the dataflow being filled with the buffered events in the source task. 
We can also clearly see that DSM takes much longer to reach a stable output rate, at $\approx 480~secs$, relative to DCR and CCR which flatten out at $\approx 320~secs$.

The latency plot, Fig~\ref{fig:up:latency}, shows 
three horizontal Red/Blue/Green lines that represent the median latency of the DAG when stable for the respective strategy. The vertical dashed lines indicate the event timestamp corresponding to the various metrics, as labeled in the caption. We can see that average latency of the DAG is high for DCR and CCR between $\approx 140-300~secs$, but returns to the steady latency beyond that. However, with DSM,  we reach this much later at $\approx 390~secs$. Similarly, the other metrics reported in Fig~\ref{fig:timing} and the analysis above can be correlated here.

\textbf{Summary.} Our results show that CCR can be used for reliable DAG migration with a recovery time of less than $27~sec$ for any DAG size, during which time we do not see any output events. Also, it can catchup with old events within $50~secs$ and the output rates stabilize within $140~secs$. These indicate a rapid completion of migration for Storm dataflows that allow it to exploit event the per-minute and per-second billing that Cloud providers are offering. This is also important for latency-sensitive streaming applications. DCR can be preferred if we need guarantees that old events before migration must be processed separately, and not interleave with new events. This may happen if the dataflow logic is being changed as part of migration. However, its drain time is sensitive to the critical path of the DAG or input event rate. 
DSM performs uniformly bad across all metrics.
We can see little different in the impact of either scaling in or scaling out. So our migration techniques can be easily adapted to enact diverse elastic scheduling scenarios for streaming applications.

\section{Related Work}
\label{sec:related}



Several papers have examined elasticity for stream processing systems. This is true not just within Clouds, also for Edge based systems like~\cite{lujic:icfec:2017} that motivates dynamic migration of the task due to edge resource outages.

\emph{Stela}~\cite{xu:ic2e:2016} built on Apache Storm does on-demand scaling, and is similar to our work. It optimizes the dataflow throughput while limiting the interruption of the running application. The paper assumes that operators are stateless in Storm, which avoids issues of state handling and consistency. It too uses \emph{convergence time}, similar to our stabilization time, as the comparison metric but does not consider message delivery guarantees or message failure due to migration into account. It uses \emph{effective throughput} as a measure to decide if an operator has to be migrated for the given resources. We do not focus on the resource allocation problem here, but only on reliable migration once the allocation is decided.

\emph{E-Storm}~\cite{liu:icpp:2017} has extended the default Redis-based state preservation in Storm to a replication-based state management that actively maintains multiple state backups on different nodes. The recovery module then restores the lost state by inter-task state transfer. This can only mask the loss of a task states in case of JVM and the host crash, but not when the DAG is migrated during scaling in/out. They claim improvements in throughput by avoiding access to a remote data store for the state.

\emph{Esc}~\cite{satzger:cloud:2011} considers events as key-value pairs. Dynamic load balancing is done by mapping keys to different VMs using hash functions. The execution model instantiates 
multiple tasks at runtime as needed. While it can dynamically adjust the Cloud resources based on the current load, it uses its custom streaming platform. We instead investigate supporting reliable elasticity within the popular Storm DSPS. 

\emph{Gedik, et al.}~\cite{gedik:tpds:2014} have proposed elastic auto-parallelization for balancing the dynamic load in case of streaming applications. They dynamically adjust the data parallelism to handle the congestion using a control algorithm. An incremental migration protocol is proposed for maintaining states while scaling, minimizing the amount of state transfer between hosts. But their evaluation does not consider message reliability during the scaling and state migration operation.

\emph{ElasticStream}~\cite{ishii:cloud:2011} uses a hybrid Cloud model for stream processing to address a lack of resources on the local stream computing cluster. They dynamically adjust the resources required to respond to dynamic rates, with the goal of minimizing the cost for using the Cloud while satisfying their QoS. The implementation on \emph{IBM's System S} is able to assign or remove computational resources dynamically. Though the system transfers data stream processing to Cloud, they do not talk about message reliabilty and state handling during the migration. This may lead to loss of in-flight messages and internal state of stateful tasks during migration. 

Our prior work~\cite{kumbhare:sc:2013} has proposed the use of \emph{alternate tasks} to allow changes to the streaming dataflow structure, and uses these to adapts to varying performance of cloud VMs. Dynamic input rates are managed by allocating resources for alternate tasks, thus making a trade-off between cost and QoS on Cloud. Our current work addresses static dataflow structures but can adapt to changes in input rates and support elastic migration. 

A related topic to elasticity of streaming platform is VM migration in the Cloud, contrasting PaaS and IaaS approaches. ~\cite{zhang:noms:2012} have proposed a VM migration algorithm, Scattered, that minimizes VM migrations in over-committed data centers. These VMs are migrated to underutilized physical hosts to balance its utilization. Voorsluys, et al.~\cite{buyya:cloudcom:2009} evaluate the effects of live VM migration on the performance of running applications. 
Others~\cite{zhao:vtdc:2007} have proposed a $3$ phase VM migration approach: suspend, copy and resume, where the  VM memory is first captured, it is suspended at the origin host and then the VM's configuration and memory state transferred to the destination host. During resume, the memory state is restored from the snapshot and then execution is resumed. This basic idea is quite similar to our CCR approach.

PaaS Cloud providers have to manage resources of customer applications to trade-off IaaS costs and QoS. 
CloudScale~\cite{shen:cloud:2011} adjusts the resources assigned to each VM in a given host using a workload predictor. When the forecasted load for a host exceeds its capacity, the VM is migrated.
\cite{casalicchio:cacc:2013} propose an algorithm that analyzes the negative impact of VM migration on an IaaS provider. An optimization problem is formalized and best solution is provided using a hill climbing search.

Proactive and preventive checkpointing has been explored for large HPC systems~\cite{bouguerra:ipdps:2013} to increase the computational efficiency during failures. Formal models have been proposed for failure prediction based on the checkpoint cost, the failure distribution, and the probability of success of the proactive action. ~\cite{naksi:ccgrid:2008} have used an incremental Checkpoint/Restart approach that tries to reduce the large memory use by switching to reliable storage for full checkpoints.




\section{Conclusions}
\label{sec:conclusions}
In this paper, we have presented dynamic migration techniques for streaming dataflows that help fast data platforms rapidly exploit the elasticity offered by Clouds. We have implemented and validated the DCR and CCR strategies that we have proposed using existing rebalance and State checkpointing mechanisms available in Apache Storm, and compared it with its native migration feature. 
Our validations using micro and application DAGs and for scaling out and in on Cloud VMs show significant benefits of both these approaches over DSM. CCR, in particular, is able to 
restore the dataflow state and behavior after migration within $50~sec$ while the default approach take over $100~secs$, and grows with the DAG size. This makes CCR beneficial for fine-grained elasticity and cost reduction on pay-as-you-go Cloud IaaS, while not compromising the performance of such fast data applications. 
The uses of this capability are plenty. We can further extend and use DAG migration for interesting problems like updating the task logic by re-wiring the DAG on the fly, task migration due to insufficient storage availability, for dynamic DAG resource updation to support certain latency requirements.


\IEEEtriggeratref{23}


\bibliographystyle{IEEEtran}
\bibliography{paper}

\end{document}